\documentstyle[12pt]{article}
\begin{document}
\title{Non-Commutative Geometry, Spin and Quarks}
\author{B.G. Sidharth$^*$\\
B.M. Birla Science Centre, Hyderabad 500 063 (India)}
\date{}
\maketitle \footnotetext{$^*$E-mail:birlasc@hd1.vsnl.net.in}
\begin{abstract}
In this paper we use considerations of non-commutative geometry to
deduce a model for QCD interactions. The model also explains
within the same theoretical framework hitherto purely
phenomenological characteristics of the quarks like their
fractional charge, mass,handedness and confinement.
\end{abstract}
\section{Introduction}
In recent years it is being realized that an important assumption
that has been uncritically taken for granted in much of twentieth
century Physics is that space time forms a differentiable
manifold. This is the case, for example in the Riemannian General
Relativity, and the Minkowski Space Time of relativistic Quantum
Mechanics and Quantum Field Theory. It is only in recent years
that space time manifolds where we cannot go down to arbitrarily
small intervals have been considered in the context of, for
example non-commutative space time models, SuperString Theory and
Quantum Gravity and similar areas \cite{r1,r2,r3,r4}.\\
Indeed it has been suggested by the author that once we discard
the conventional space time geometries and consider a non-
commutative space time then it is possible to reconcile the
hitherto irreconcible pillars of the twentieth century, namely
General Relativity and Quantum Theory \cite{r5,r6,r7,r8}. We will
now argue that from the same non- commutativity, it is possible to
also understand strong interactions and the quark picture.
\section{Strong Interactions}
Let us introduce the effect of a non-commutative space time into
the usual metric, at some scale $(l,\tau)$
\begin{equation}
ds^2 = g_{\mu \nu} dx^\mu dx^\nu\label{e1}
\end{equation}
If we split up the product $dx^\mu dx^\nu$ into symmetric and non
symmetric parts, (\ref{e1}) will become
\begin{equation}
g_{\mu \nu} = \eta_{\mu \nu} + h_{\mu \nu}\label{e2}
\end{equation}
In (\ref{e2}), the first term on the right side represents the
usual metric, while the second term represents the effect of non-
commutativity of the space time coordinates viz.,
\begin{equation}
[x,y] = 0(l^2), [x,p_x] = \imath \hbar [1 + l/\hbar )^2
p^2_x]etc.\label{e3}
\end{equation}
Equation (\ref{e3}) can be deduced from relations that were worked
out by Snyder \cite{r9,r10}, and subsequently by several other
scholars over the past five decades and more. Incidentally if in
(\ref{e3}), we specialize to the Compton scale, i.e. $l$ denotes
the Compton wavelength (and similarly, $\tau$, the Compton time),
then, we can infact deduce the Dirac equation of the electron
\cite{r11,r12}. In other words the non- commutativity is an $\sim
0(l^2)$ effect. Starting from (\ref{e2}), it has been shown in
detail that we can deduce the gravitational field equations
\begin{equation}
\Box \phi^{\mu \nu} = - kT^{\mu \nu}\label{e4}
\end{equation}
where
\begin{equation}
\phi^{\mu \nu} = h^{\mu \nu} - \frac{1}{2} \eta^{\mu \nu}
h\label{e5}
\end{equation}
(Cf.\cite{r5,r6}for details).\\
Equations (\ref{e4}) and (\ref{e5}) represent the linearized
equations of General Relativity \cite{r13}.\\
Starting from equation (\ref{e4}), we have, as is well known, in
suitable units,
\begin{equation}
g_{\mu \nu} = \eta_{\mu \nu} + h_{\mu \nu}, h_{\mu \nu} = \int
\frac{4T_{\mu \nu}(t-|\vec x - \vec x |, \vec x')}{|\vec x - \vec
x' |} d^3 x'\label{e6}
\end{equation}
It may be mentioned that in (\ref{e6}), velocities comparable to
that of light are allowed, and at the same time the stresses
$T^{\imath j}$ and momentum densities $T^{0j}$ can be comparable
to the energy momentum density $T^{00}$.\\
It is well known that when
$$| \vec x' | < < | \vec x |,$$
we have the equations
\begin{equation}
m = \int T^{00} d^3 x\label{e7}
\end{equation}
\begin{equation}
S_k = \int \epsilon_{klm}x^lT^{m0}d^3 x\label{e8}
\end{equation}
where $m$ is the mass and $S_k$ is the angular momentum. In any case, as is well
known, (\ref{e6}) leads to the gravitational potential \cite{r13}. It may be mentioned
that integrals and derivatives within the non-commutative geometry are approximate,
because, in any case, the effects are, as pointed out, $\sim 0(l^2)$.\\
We consider the integrals in (\ref{e7}) and (\ref{e8}) in a region
bounded by the Compton wavelength, because in any case, the
density of the particle vanishes outside this region, and hence,
so also the energy momentum density $T^{\mu \nu}$. Remembering
that the velocity at the Compton wavelength equals $c$, the
velocity of light, we can easily deduce from (\ref{e8}) that the
angular momentum $S_k$ is given by,
$$S_k = \frac{h}{2},$$
that is the Quantum Mechanical spin half. (Cf.\cite{r12} for details]).\\
Indeed from an alternative viewpoint it has been shown that the
non-commutative
relations (\ref{e3}) imply spin and conversely \cite{r14}.\\
Let us now consider the case when
$$| \vec x' | \sim | \vec x |$$
Then we have from (\ref{e6}), expanding in a Taylor series about
$t$,
$$h_{\mu \nu} = 4 \int \frac{T_{\mu \nu}(t,\vec x)}{|\vec x - \vec x'|}
d^3 x' + (\mbox{terms independent of}\vec x) +2$$
\begin{equation}
\int \frac{d^2}{dt^2}T_{\mu \nu} (t,\vec x), |\vec x - \vec
x'|d^3x' + 0 (|\vec x - \vec x' |^2)\label{e9}
\end{equation}
To proceed further, we will need the relation,
\begin{equation}
|\frac{du_\nu}{dt}| = |u_\nu |\omega\label{e10}
\end{equation}
where $\omega$ the frequency is given by,
$$\omega = \frac{|u_\nu}{R} = \frac{2mc^2}{\hbar}$$
Equation (\ref{e10}) can be derived in a simple way as follows:
The non-commutative relations (\ref{e3}) imply
$$y \equiv \tilde {h} p_x \left(\tilde {h} = \frac{H}{h} , H = 0(l^2)\right)$$
(Cf. ref. \cite{r7,r8}).\\
Whence we have
$$\dot {u_\mu} = \frac{1}{m} (\dot {p_\mu}) = \frac{1}{m}
\frac{h}{H} (\dot x_\nu) = \frac{1}{m} \frac{h}{H} u_\nu ,$$ which
leads to (\ref{e10}). Interestingly, this is also true in
the theory of the Dirac equation itself.\\
Using (\ref{e10}), we have,
$$\frac{d}{dt} T^{\mu \nu} = \rho u^v \frac{du^\mu}{dt} + \rho u^\mu
\frac{du^\nu}{dt} = 2 \rho u^\mu u^\nu \omega ,$$
so that,
$$\frac{d^2}{dt^2} T^{\mu \nu} = 4 \rho u^\mu u^\nu \omega^2 = 4 \omega^2 T^{\mu \nu}$$
where $\omega$ is given in (\ref{e10}). Substitution in (\ref{e9})
now gives,
\begin{equation}
h_{\mu \nu} \approx \frac{\beta M}{r} + 8\beta M
(\frac{Mc^2}{\hbar})^2 \cdot r\label{e11}
\end{equation}
$\beta$ being a constant.\\
This resembles the QCD quark potential \cite{r15}. In any case
these considerations suggest that we can get different
interactions at different distances or scales in a
unified picture, which can approximately atleast represent quarks also.\\
We can further refine this argument. For this we will use the
following relation already deduced (Cf.\cite{r12} and references
therein for details):
$$A_\sigma = \frac{1}{2} \left(\eta^{\mu \nu} h_{\mu \nu}\right),_\sigma ,$$
It was then shown that the electromagnetic potential is given by,
$$\frac{e^2}{r} = A_0 \approx \frac{2c\hbar}{r}\int \eta^{\mu \nu}
\frac{d}{d\tau} T_{\mu \nu} d^3x' = \frac{2c\hbar}{r} \int
\eta^{\imath j}\frac{d}{d\tau} T_{\imath j} d^3x',$$
\begin{equation}
= 2c\hbar (\frac{mc^2}{\hbar}) \int \eta^{\imath j}
\frac{T_{\imath j}}{r} d^3 x',\label{e12}
\end{equation}
outside the Compton wavelength.\\
As we approach the Compton wavelength however, we have to use
equation (\ref{e9}), which after a division by $m$, the mass of
the particle to be identified with the quark, and taking $\hbar =
1 = c$ to correspond to the usual theory, goes over to,
\begin{equation}
- \frac{\alpha}{r} + \frac{\beta m_e}{l^2} r\label{e13}
\end{equation}
which is essentially (\ref{e11}).\\
In (\ref{e13}) $\alpha \sim 1 \mbox{and} \beta \sim \frac{1}{m}$
and $m_e$ is the electron mass.
This is the QCD potential with both the Coulumbic and confining parts (Cf.ref.\cite{r15}).\\
We now observe that the usual three dimensionality of space, as
pointed out by Wheeler \cite{r13} arises due to the double
connectivity or spinorial behaviour of Fermions, which takes place
$\underline{outside}$ the Compton wavelength due to the fact that
as has been seen elsewhere, while it is the negative energy
components of the Dirac four-spinor which dominate inside, it is
the positive energy components which predominate outside
(cf.ref.\cite{r5,r12} for details). Such a three dimensionality
can also be deduced using Penrose's spin network theory
\cite{r16}. Interestingly, if we consider the Dirac equation in
two (or one dimension) \cite{r17,r18}, we encounter handedness and
the absence of an invariant mass - features which in the light of
our considerations arise $\underline{at}$ the Compton wavelength.
As we approach the Compton wavelength, we encounter mostly the
negative energy component and the above double connectivity and
therefore three dimensionality disappear: We have two or less
dimensions. Indeed even in the purely classical case of a
collection of relativistic particles, the various centres of mass
form a two dimensional disk\cite{r19}. Such a conclusion has been
drawn alternatively at
very small scales (cf.\cite{r20,r21}).\\
This leads to the following circumstance: We have to consider two
and one spatial dimensions. We now use the fact that as is well
known \cite{r22} for each dimension the $T_{\imath j}$ in
(\ref{e9}) or (\ref{e12}) is given by $(1/3)\epsilon$, where
$\epsilon$ is the energy density. In this case it follows from
(\ref{e12}) that the particle would have the charge $(2/3)e$ or
$(1/3)e$, in two or one dimensions. Incidentally, this provides an
explanation for the remarkable and well known fact that one third
of charge appears to be concentrated in a core of the size of the
order of the proton Compton wavelength as was
experimentally well established \cite{r23}.\\
Using the fact that at the Compton wavelength the charge becomes
$e/3$, and $d^3r \to l^2 dr$ owing to the single dimensionality in
equation (\ref{e12}), and also using equation (\ref{e10}) in
(\ref{e9}) we get
$$\frac{1}{9\times 137 r} \sim m_e \cdot l^2 \int \frac{T}{r}
dr \quad \mbox{or} \quad  \int \frac{T}{r} \sim \frac{1}{r \times
10^3 m_e} \cdot m^2 \sim \frac{m}{r}$$ where the last step follows
from a comparison with the Coloumb part of (\ref{e11}) or
(\ref{e13}). Whence
\begin{equation}
m \sim 10^3 m_e\label{e14}
\end{equation}
where $m$ is the quark mass, remembering that in the above natural units, $l = \frac{1}{m}$.\\
Equation (\ref{e14}) is ofcourse correct. Infact using the quark mass given in (\ref{e14}) in
(\ref{e11}) or (\ref{e13}), it is easy to see that the ratio of the coefficients of
the confining and Coulumbic parts is $\sim (Gev)^2$ which is also true \cite{r15}.\\
Thus the quark and QCD identification is complete, and also at the
same time a theoretical rationale for the mass, fractional charges
and handedness of the quark has now been obtained. As noted by
Salam,
these were hitherto not explained theoretically \cite{r24}.\\
This would also automatically imply that these fractionally
charged particles cannot be observed individually, as they by
their very nature appear when confined to dimensions of the order
of their Compton wavelength. This is expressed by the confining
part of the QCD potential (\ref{e11}) or (\ref{e13}).
\section{Remarks}
We would like to reemphasize that our model gives the mass (order
of magnitude), the fractional charge, the handedness and the
confinement feature of quarks correctly. Experimentally, a charge
$e/3$ has been observed within the quark Compton wavelength,
as pointed out. The correct phenomenological QCD potential also is obtained.\\
A final remark: In the above considerations, we had specialized to
the Compton wavelength. We can see from an alternative and simple
point of view, how the compton scale emerges. For this we use the
fact that in the Dirac theory of the electron, we have
(Cf.ref.\cite{r14,r25})
$$\hat {p} (\approx \frac{\hbar}{\hat{x}}) = \frac{E^2}{\hbar c^2} \hat{x}$$
whence
$$\hat{x} \approx \frac{\hbar}{mc},$$
where $\hat{x}, \hat{p}$ arise due to well known non-Hermitian
effects. That is we recover the Compton scale, within which
non-commutative and non-Hermitian effects come into play.

\end{document}